# The Great Aurora of January 1770 observed in Spain

Víctor M.S. Carrasco[1], Enric Aragonès[2], Jorge Ordaz[3], José M. Vaquero[4,5]

[1]Departamento de Física, Universidad de Extremadura, Badajoz 06071, Spain
[2]Societat Catalana d'Història de la Ciència i de la Tècnica, Barcelona 08011, Spain
[3]Instituto Feijoo del Siglo XVIII, Universidad de Oviedo, Oviedo 33005, Spain
[4]Departamento de Física, Universidad de Extremadura, Mérida 06800, Spain
[5]Instituto Universitario de Investigación del Agua, Cambio Climático y Sostenibilidad (IACYS), Universidad de Extremadura, 06006 Badajoz, Spain

*Correspondence to*: Víctor M.S. Carrasco (vmscarrasco@unex.es)

**Abstract.** An analysis is made of the records made by Spanish observers of a notable aurora on 18 January 1770 in order to study the characteristics of this event. The records indicate that the phenomenon was observed in both continental and insular territories of Spain, in particular at San Cristóbal de la Laguna, Cádiz, Córdoba, Badajoz, Valencia, Castellón, Madrid, Barcelona, and Gerri de la Sal. The most equatorward observational site was San Cristóbal de la Laguna (28.48° N, 16.32° W) in the Canary Islands. In general, the descriptions put its duration from sunset to midnight, but the observers from Córdoba and Madrid report the aurora as being visible during the last hours of the night, and it was even observed the following day at Castellón. All the observers described the aurora as red in colour, while white and ash colours were also reported at Córdoba and Gerri de la Sal. The brightness and shape of auroral display changed over time. Calculations of the geomagnetic latitudes of the observation locations gave San Cristóbal de la Laguna as the southernmost (26° N) to Gerri de la Sal the northernmost (35° N), and indicate this aurora was observed over a wide range of abnormally low latitudes for such a phenomenon. Solar activity around the event was high, with the astronomer Horrebow registering 10 sunspot groups on that date (18 January 1770).

## 1 Introduction

An aurora borealis is a natural phenomenon that has impressed different civilizations since ancient times, generating both admiration and fear (Eather, 1980). It originates when the solar wind interacts with the Earth's magnetosphere. This interaction causes the release of particles already trapped near Earth, which then collide with oxygen and nitrogen in the upper atmosphere producing striking lights in the sky. Thus, aurorae represent one of the several manifestations of solar activity, and records of them are of great interest since they can be taken as a proxy with which to study the behaviour of past solar activity (Siscoe, 1980; Silverman, 1992; Willis *et al*., 2009; Vázquez *et al*., 2016).

The appearance of an aurora borealis at low magnetic latitudes is infrequent. It is generally associated with intense geomagnetic storms, but an aurora has sometimes been observed at low latitudes when geomagnetic activity was weak or moderate (Silverman, 2003; Willis *et al*., 2007). The magnetic latitude in Spain is relatively low, especially in the Canary

Island. For this reason, Spanish auroral reports can be a good detector of large geomagnetic storm. For example, during the great Carrington storm, auroral displays were also seen in Spain (Vaquero *et al.*, 2008). In today's world, geomagnetic storms can cause major problems due to our dependence on technology which is vulnerable to electromagnetic perturbations. Some potential consequences of these events are disturbances in communication systems, power blackouts, and permanent damage to transformers (Pulkkinen, 2007). It is therefore essential to understand and predict the behaviour of these phenomena.

In the past, several aurorae borealis have been observed in Spain and Portugal. The first Spanish catalogue was made in the 19th century (Rico Sinobas, 1855; Vaquero et al., 2003). Vaquero & Trigo (2005) and Vaquero *et al.* (2010) report systematic aurora observations in Lisbon compiled by Jacob Prætorius and Henrique Schulze during the 18th century, and in Barcelona by Francisco Salvá during the period 1780-1825. Other, non-systematic, auroral observations from Spain and Portugal have been published by Vaquero *et al.* (2003) and Carrasco *et al.* (2017). Aragonès & Ordaz (2010) compiled 80 auroral display observed in Spain and Portugal during the 18th century. This catalogue includes original descriptions of the great aurora that occurred in January 1770. Schröder (2010) discussed the development of this aurora, showing that it was observed at middle and low latitudes based on Fritz (1873) and Angot (1896). It was also used in Vázquez *et al.* (2006) to show that the open magnetic field better describes middle and low latitude auroral occurrences.

The objective of the present work is to analyse several records made by Spanish observers concerning an aurora that occurred on 18 January 1770, records contained in the catalogue published by Aragonès & Ordaz (2010). In Sec. 2, we provide the locations where this auroral display was observed. Section 3 summarizes the original aurora records given in Aragonès & Ordaz (2010). Section 4 analyses and discusses the results, and Sec. 5 presents the main conclusions.

**2 Geographic Distribution of the Observations**

Aragonès & Ordaz (2010) compiled a catalogue with 80 auroral display observed in the Iberian Peninsula and the Balearic and Canary Islands during the 18th century. In particular, this catalogue contains information about a notable aurora that occurred on 18 January 1770, including descriptions and the observational sites. This aurora was also sighted at other, more northern, European locations such as London (51° 30' N, 0° 7' W) and Berlin (52° 31' N, 13° 23' E), and at others of similar latitude to those in Spain such as Rome (41° 53' N, 12° 30' E) and Naples (40° 50' N, 14° 15' E) (Fritz, 1873; Angot, 1896; Schröder, 2010).

Several observers located at different places in Spain registered this exceptional aurora (Fig. 1). The northernmost site was Gerri de la Sal (42.32° N, 1.07° E), and the southernmost San Cristóbal de la Laguna (28.48° N, 16.32° W). The other sites were Cádiz (36.52° N, 6.28° W), Córdoba (37.88° N, 4.77° W), Badajoz (38.88° N, 6.97° W), Valencia (39.47° N, 0.38° W), Castellón (39.97° N, 0.05° W), Madrid (40.42° N, 3.69° W), and Barcelona (41.38° N, 2.18° E).

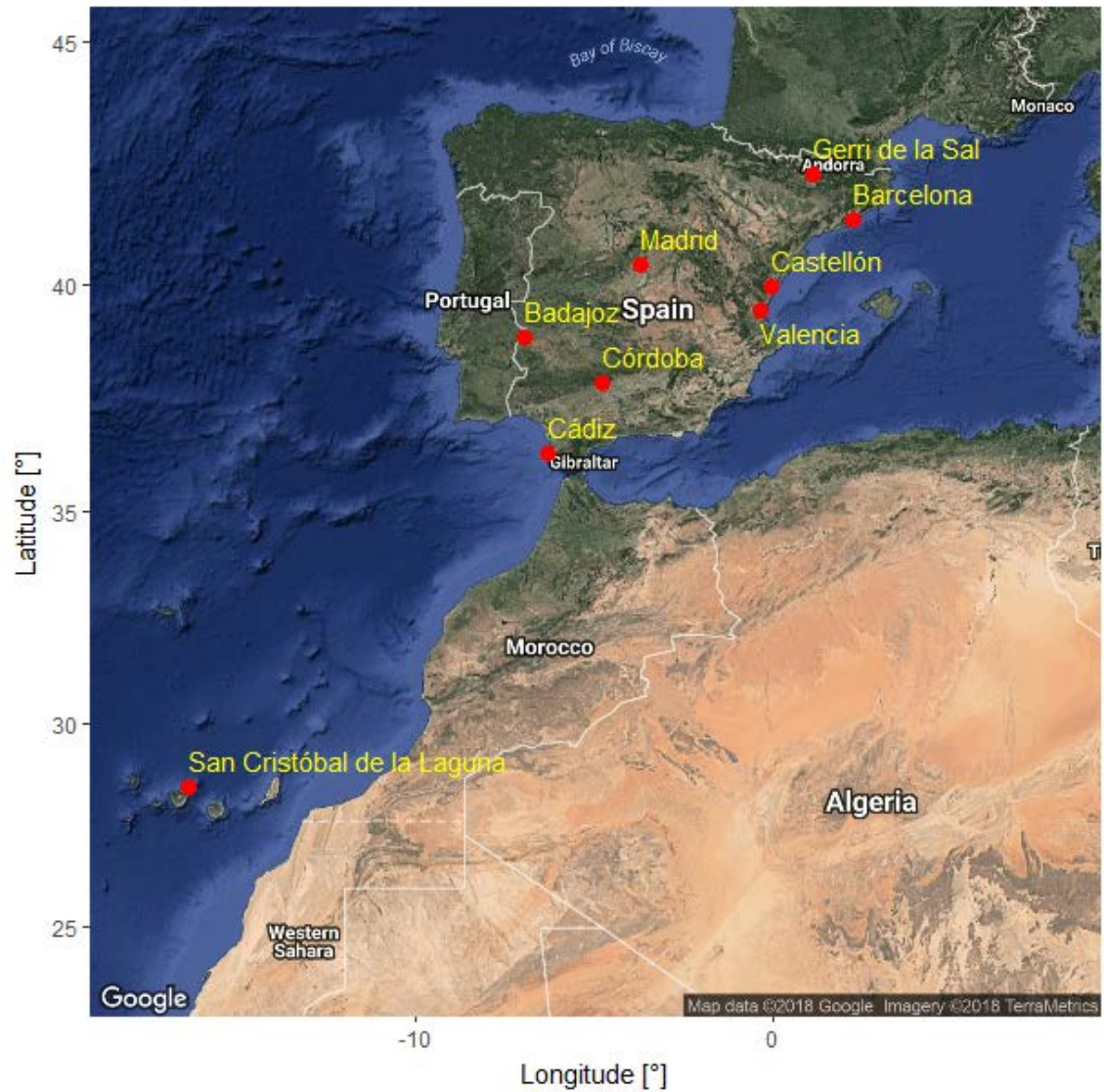

**Figure 1:** Locations (red dots with yellow letters) in a modern map where a noticeable aurora was registered in Spain on 18 January 1770.

**3 Descriptions of the Aurora Records**

Aragonès & Ordaz (2010) transcribe the original descriptions made by several Spanish observers about the aurora observed on 18 January 1770. Here, we shall present a summary of those records taken from Aragonès & Ordaz (2010), highlighting the most notable characteristics registered at each observation site.

The aurora observed in Córdoba was sighted on 18 January 1770 from 17:30 (local time) to midnight, but some of the countryfolk ("hombres de campo", i.e., men living and working outside the city itself) affirmed that the aurora remained until sunrise on the following day. According to the Professor of Arts of the Colegio de la Asunción, the aurora developed in the northern sky, and was changing in shape, brightness, and colour (Nipho, 1770). Figure 2 is the header of the article in which the description of the observation of the aurora in Córdoba on 18 January 1770 was published. Original text – "[…] dia 18. de enero, como a las cinco y media de la tarde […] Comenzose á dexar ver por el Nordeste, y poco despues se fue manifestando por el Norte haciendo columnas, ó rafagas de bastante diametro, y que se dividian unas de otras […] Es cosa digna de notar que esta Aurora era en su mayor parte materia inflamada, y que solo por abaxo blanqueaba un poco […] á las once y media yá nada se descubria quedó la noche tan obscura, como qualesquiera otra sin Luna […] aseguran algunos hombres del campo haberse conservado esta aurora yá mas, yá menos viva hasta el amanecer"; translation – "On 18 January 1770 around 17:30 […] It started to be seen in the northeast, and after it manifested in the north as columns or bursts with great diameter and divided from each other […] It is noteworthy that this Aurora was for the most part inflamed matter, and it only blanched a little at the bottom […] At 23:30, nothing was discoverable, the night was as dark as any other without a moon, […] some countryfolk [men from outside the city] declared that the aurora was preserved more or less alive until sunrise."

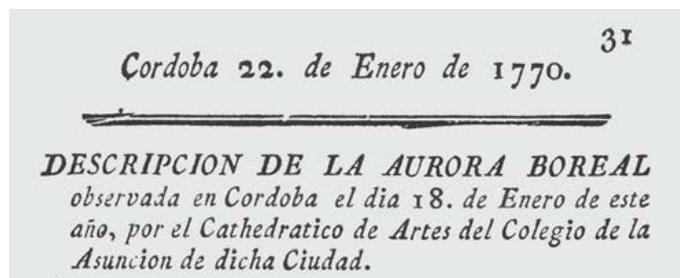

**Figure 2:** Header of the article about the observation of aurora made by the Professor of Arts of the Colegio de la Asunción in Córdoba on 18 January 1770 [Source: Nipho, 1770].

In Madrid, the aurora was observed at nightfall to the north. It presented several shapes, a red colour, and a brightness that rose and fell over time. The sky was clear and the aurora was visible all night (Subirás y Barra, 1770). Original text – "Jueves último dia 18 al anochecer se dexó ver una fuerte Aurora Boreal […] la Aurora Boreal consistía entonces en una faxa como de 20° de ancho de un color rojo muy debil, y casi uniforme en toda su longitud, de figura de una faxa esferica ó semicorona que salia se de hacia al este, y pasando por el cenit de Madrid se terminaba hacia el Nor-nordoeste […] el color rojo se iba aumentando, y disminuyendo por poco tiempo á trechos […] A las 6 de la noche refieren era semejante a lo que expreso de las 10 ½"; translation – "Thursday 18 at dusk, a strong Aurora Borealis was seen, […] the Aurora Borealis consisted of a girdle about 20° wide with a very weak red colour, and almost uniform over all its length, with a figure of a spherical girdle or semi-corona from the east and, passing through the zenith of Madrid, ending up towards the north-northwest […] the red colour intensified and faded over short time intervals […] At 06:00 [19 January 1770], the aurora was similar to the description for 22:30 [18 January 1770]".

The description provided by an anonymous observer in Castellon indicated that the aurora was sighted in the northern sector of the sky, that it was red in colour, and that it remained in the same state from 19:30 (local time), when it started to be seen, until midnight (Anonymous, 1770a). This observer reported that the aurora was also seen the next day (19 January) at around 20:00 (local time). Original text – "A las 7 ½ me avisaron avia una[s] nubes coloradas subí a la torre de casa y vi unas Auroras Boreales una al Nordoestoest […] y la otra menos grande a Nord de un color rojo obscuro melancolico […] A las onse de la noche volví a verla y encontre de 4 a 5 auroras boreales de NE a NO formando nubes mas altas que anchas […] me han asegurado que a las 12 permanesian en el mismo estado"; translation – "At 19:30, I was alerted that there were reddish clouds, I went up to the tower of my house and I saw some aurorae borealis, one to the west-northwest […] and the other, less large, to the north with a dark melancholic red colour […] At 23:00, I went back to see it and found from 4 to 5 aurorae borealis from northeast to northwest forming clouds higher than wide […] other people have assured me that they remained in the same state at 24:00."

Another anonymous observer at Gerri de la Sal noted that "a tower of fire" appeared in the north of the sky at 18:30 (local time), and it later extended from east to west showing different shapes and white and ash colours. It disappeared at midnight (Anonymous, 1770b). Original text – "[…] a cosa de las sis, y mitja aparegue en est emisferi en la part de tramontana una com torre de foch, que feia unas vias de color de llet y a cosa de mitja hora se dividí, y extengué ab una gran nuvolada, també inflamada, per les parts de Orient, y Ponent, dexant la part de mitxdia libre estant tot lo cel seré pero de color de cendra […] y à las once horas […] se desvanesqué tot"; translation – "[…] At 18:30, a tower of fire appeared in the hemisphere in the tramontane part [north], which made some lines of milk colour and half an hour later split, and spread out with a large cloud, also inflamed, in the parts of East and West with the part of midday [south] free, being all the sky clear but with an ash colour […] and at 23:00 […] everything disappeared".

Aragonès & Ordaz (2010) also provide the original description made by the historian and naturalist José de Viera y Clavijo (Fig. 3) at San Cristóbal de la Laguna (Guerra y Peña, 1951). This observer indicated that the red glow of the aurora spread over all the northern part with a very bright light from one hour after twilight to midnight. Original text – "Algo más de una hora después del crepúsculo, se extendió por la ciudad el rumor de que los montes Taganana quizás estuvieran ardiendo […] Salí a observar el fuego, pero para mi sorpresa me encontré ante una verdadera aurora boreal […] y la aparición en forma de llama de color rojo sangre se extendía por todo el norte desde el este hasta unos pocos grados más allá del oeste, con una luz muy brillante"; translation – "Something more than an hour after twilight, there spread throughout the city a rumour that the Taganana hills might be burning […] I went out to observe the fire, but, to my surprise, I found myself before a true aurora borealis […] and the appearance [of the aurora] in the shape of a flame of a blood-red colour extended over all the north from east to a few degrees beyond west, with a very bright light." A complete English translation of that record can be consulted in Vázquez *et al.* (2006).

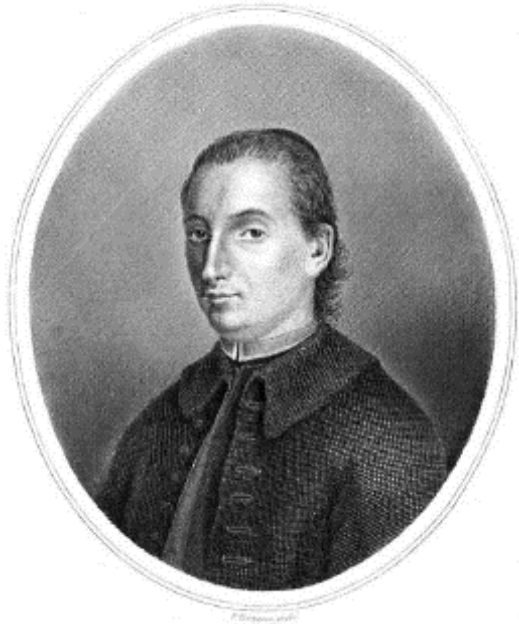

**Figure 3:** Portrait of José de Viera y Clavijo (1731-1813), observer of the aurora from San Cristóbal de la Laguna (Canary Islands) on 18 January 1770 [Source: http://bdh-rd.bne.es/viewer.vm?id=0000042056&page=1, Biblioteca Nacional Española].

For the case of Barcelona, Aragonès & Ordaz (2010) found that an aurora was observed from 19:00 to midnight as a cloud of fire in the sky split into four parts with a great radiance (Sagarriga, 1907, p. 57). Aragonès & Ordaz (2010) presented the description of the aurora that was made by Richard (1771), noting that the same variations of the phenomenon observed in Genoa were also seen in Cádiz. Moreover, Aragonès & Ordaz (2010) reported that this aurora was also recorded in Valencia as "one of the most complete aurorae observed" (Pastor Fuster, 1830, p. 471), and by the priest Leonardo Hernández Tolosa in Badajoz where several people observed how the northern sector appeared reddened from sunset until two hours after midnight (Hernández Tolosa, 1992, p. 198; Vaquero, 2001).

**4 Analysis and Discussion of Results**

The aurora that occurred on 18 January 1770 was observed over a wide zone of latitudes that were abnormally low given the typical behaviour of this kind of phenomenon. In Spain, the aurora was recorded from San Cristóbal de la Laguna (28.5° N) to Gerri de la Sal (42.3° N). It was also observed over all of the rest of Europe (Richard, 1771; Fritz, 1873; Angot, 1896).
The likelihood of seeing an aurora is generally greater at higher geomagnetic latitudes. To investigate this condition, we computed the geomagnetic latitudes of the locations considered in the present work at which aurorae were registered on 18 January 1770. To this end, we computed the geomagnetic latitudes using the dipole model of the "gufm1" geomagnetic model (Jackson *et al*., 2000) and the Web application of the National Geophysical Data Center (https://ngdc.noaa.gov/geomag-web/#igrfwmm). This model is based on a massive compilation of historical observations of the magnetic field. Before 1800, more than 83000 magnetic declination observations were recorded at more than 64000 locations. This model provides a high accuracy, for example, declination and inclination results are typically within 30 minutes of arc. We then calculated the geomagnetic latitudes for each location (Chapman & Bartels, 1951; Udías & Mezcua, 1997). The resulting geomagnetic latitudes for the places considered here are: (i) San Cristóbal de la Laguna: 36.0° N; (ii) Cádiz: 42.6° N; (iii) Córdoba: 43.7° N; (iv) Valencia: 44.6° N; (v) Castellón: 45.0° N; (vi) Badajoz: 45.0° N; (vii) Barcelona: 46.0° N; (viii) Madrid: 46.0° N; and (ix) Gerri de la Sal: 47.1° N.

According to all the records analysed here, the aurora was observed in Spain for several hours (approximately from sunset to midnight) on 18 January 1770. However, some observers reported a longer duration. Thus, some people from country areas around Córdoba stated that the aurora was observable until sunrise of the following day (19 January). The observer at Madrid also reported that the aurora had a similar appearance at 06:00 on 19 January to the one it had at 22:30 in the previous evening. Moreover, the anonymous observer at Castellón indicated that he again saw the phenomenon at 20:00 (local time) on 19 January 1770 [Anonymous (1770a) - original text: "El dia 19 se dexo ver el fenomeno a las 8 de la noche esto es a la parte de NO y el del N no se dexo ver."; English translation: "On 19<sup>th</sup> [January], the phenomenon was seen at 8 p.m. [local time] in the northwest but not in the north"], and the Madrid observer reported a red mist or fog at dusk on 19 January [Subirás y Barra (1770) - original text: "El [día] siguiente amaneció con una niebla muy delgada que pareció disiparse á las 8, pero toda la mañana estuvo la luz del sol algo remisa. Al anochecer compareció la misma niebla toda roja, la qual se disipo luego despues de haber anochecido"; English translation: "The next [day] dawned with a very thin mist that seemed to dissipate at 8 a.m., but all morning the sunlight was somewhat reluctant. The same red fog appeared at dusk, which dissipated later after nightfall"]. All the observers indicated that the aurora developed in the northern part of the sky, and that both its brightness and its shape were changing over time. All the observers reported great brightness. The specific descriptions of the aurora varied from one another: it was recorded as a blood-red flame by José de Viera y Clavijo (San Cristóbal de la Laguna); as columns of fire, clouds, and arcs in Córdoba; as a red girdle, semi-corona, small clouds, and red mass with rays of exhalation ("effluvium") in Madrid; as red clouds with rays of light and vapour in Castellón; and as clouds, radiance, and a band and tower of fire with milky lanes in Gerri de la Sal. All the records noted the aurora as being red in colour, although in Córdoba and Gerri de la Sal the appearance of white and ash colours in the sky was also reported. We highlight that, on that day, the Moon was in the last-quarter phase and below the horizon during the main auroral event (from sunset to midnight). However, auroral display was still visible even after the moonrise in some observational sites such as Córdoba and Madrid. A summary of all these descriptions made by the Spanish observers for the auroral display on 18 January 1770 is presented in Table 1.

**Table 1.** Summary of the descriptions made by the Spanish observers for the auroral display on 18 January 1770.

| PLACE | GEOGRAPHICAL LATITUDE | GEOMAGNETIC LATITUDE | DURATION | DIRECTION | COLOR |
|---|---|---|---|---|---|
| La Laguna | 28.48° N | 26.4° N | From sunset | From east to west (northern side) | Red |
| Cádiz | 36.52° N | 31.4° N | From afternoon to midnight | From northeast to west | Red and white |
| Córdoba | 37.88° N | 32.3° N | From afternoon to sunrise | Northeast and North | Inflamed matter and blanched |
| Badajoz | 38.88° N | 33.8° N | From sunset to midnight | North | Red |
| Valencia | 39.47° N | 32.7° N | - | - | - |
| Castellón | 39.97° N | 33.1° N | From afternoon to next day | West-Northwest, north | Red |
| Madrid | 40.42° N | 34.4° N | All the night | Northeast-North | Red |
| Barcelona | 41.38° N | 33.9° N | From afternoon to midnight | - | Fire color |
| Gerri de la Sal | 42.32° N | 35.1° N | From afternoon to midnight | North and from east to west | Fire color, white and ash |

In order to contextualize the solar activity level around 18 January 1770, Fig. 4 shows the yearly Sunspot Number (version 2, http://www.sidc.be/silso/datafiles) from 1700 to 1825 (top panel), and the sunspot group counts from December 1769 to February 1770 inclusive (bottom panel), data taken from the revised collection published by Vaquero *et al.* (2016). According to the Sunspot Number Index, the year when this aurora event occurred (1770) was one year after the solar maximum of Solar Cycle 2, although the actual values of the index are very similar for these two years: 176.8 for 1769 and 168.0 for 1770. Around the date of the aurora, from December 1769 to February 1770, sunspot observations are known to have been made by three astronomers (Vaquero *et al.*, 2016): Lalande, Horrebow, and Staudach. The solar activity declined from the beginning of December to January. It then rose to reach 11 group counts on 19 January 1770, the maximum value for that 3-month period which it also reached on 7 February 1770. It is evident therefore that the solar activity level was high when the great aurora was observed on 18 January 1770. According to Horrebow's records, the group count on that day as well as the day before was 10, whereas Staudach registered 6 sunspot groups on 18 January 1770. This significant difference in the sunspot counts between Horrebow and Staudach could be due to the use of different telescopes and methodologies to observe sunspots. We would note that, approximately nine solar rotations later, around 18 September 1770, another large geomagnetic storm occurred (Aragonès and Ordaz, 2010; Hayakawa *et al.*, 2017). The number of sunspot groups that Horrebow observed on the solar disc for that day was also 10 (Vaquero *et al.*, 2016). That event has been extensively discussed by Hayakawa *et al.* (2017) who report records of the aurora observations for that day in locations with geomagnetic latitudes below 30°. These facts suggest a recurrent high geomagnetic activity in 1770.

**5 Conclusions**

A catalogue with 80 aurorae observed in Spain and Portugal during the 18th century has been compiled by Aragonès & Ordaz (2010). It includes records of a great aurora on 18 January 1770 registered in several Spanish locations: San Cristóbal de la Laguna (28.48° N, 16.32° W), Cádiz (36.52° N, 6.28° W), Córdoba (37.88° N, 4.77° W), Badajoz (38.88° N, 6.97° W), Valencia (39.47° N, 0.38° W), Castellón (39.97° N, 0.05° W), Madrid (40.42° N, 3.69° W), Barcelona (41.38° N, 2.18° E), and Gerri de la Sal (42.32° N, 1.07° E). This aurora was observed all over Europe (Richard, 1771; Fritz, 1873; Angot, 1896). We have presented a summary of the main characteristics of these records. In general, in Spain this aurora was observed from sunset to midnight. But the observers in Córdoba and Madrid reported sighting it throughout that night, and it was seen again on the next day (19 January 1770) in Castellón, and a red fog or mist was reported by the Madrid observer also on that day. The aurora was seen in the northern part of the sky according to all the records, and it was red in colour although white and ash colours were also registered in Córdoba and Gerri de la Sal. All the records describe it as being of great brightness even though this was changing over time. The aurora presented several shapes during the event, and was described differently at each place. For example, it was recorded as a blood-red flame in San Cristóbal de la Laguna, and as red columns, clouds, or a girdle in Córdoba, Madrid, Castellón, Barcelona, and Gerri de la Sal. Other terms used to define its forms were arcs, rays, vapour, and radiance. We noted that the Moon was in the last-quarter phase and below the horizon during the main hours of the event (from sunset to midnight).

We computed the geomagnetic latitudes for the locations where the records were made. These ranged from 26.4° N (San Cristóbal de la Laguna) to 35.1° N (Gerri de la Sal), showing that this aurora was observable over a broad zone of abnormally low latitudes. According to the Sunspot Number Index (version 2), this auroral event occurred one year after the maximum of Solar Cycle 2, although the Sunspot Number values of the maximum (176.8) and of 1770 (168.0) were similar. Furthermore, in the revised collection published by Vaquero *et al.* (2016), one can see that the solar activity level presented an increase around 18 January 1770. For that day, there are sunspot records made by two astronomers: Horrebow registered 10 sunspot groups on the solar disc, and Staudach 6. For the previous day (17 January 1770), 10 sunspot groups were also recorded by Horrebow.

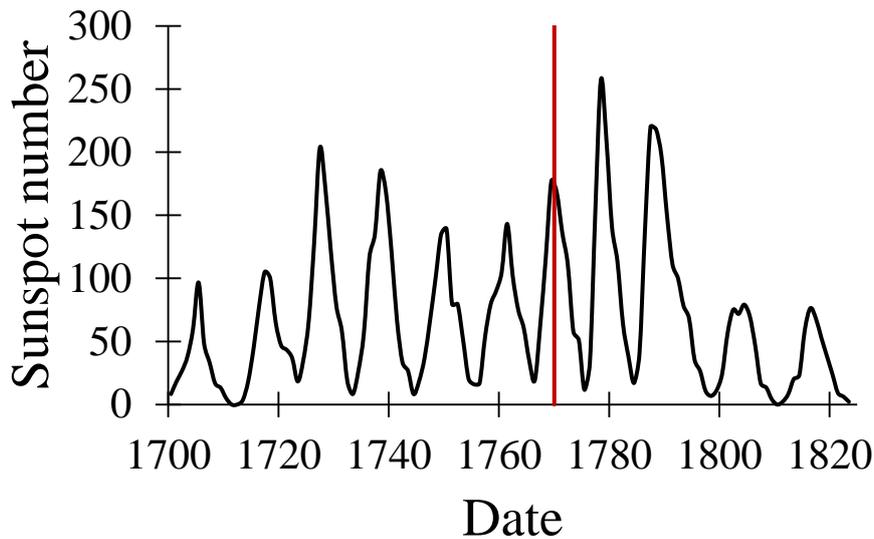

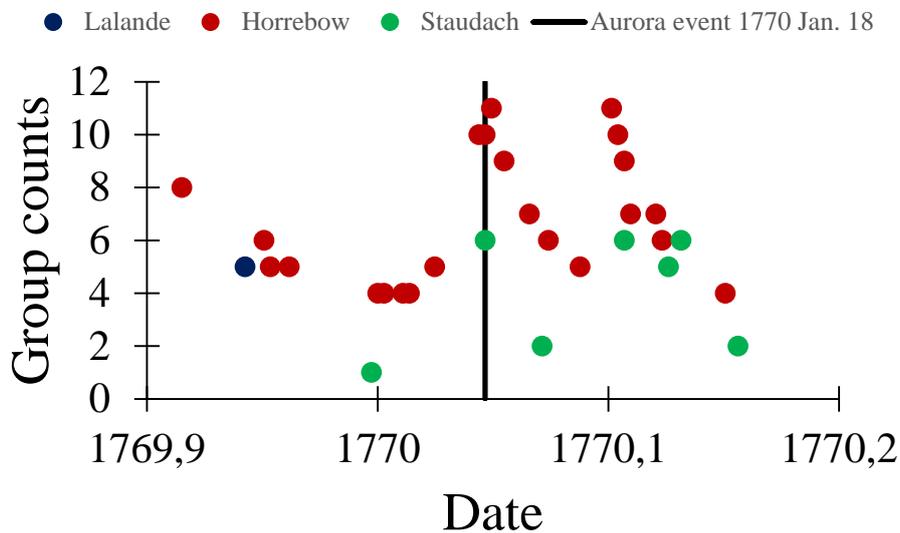

**Figure 4:** (Top panel) Yearly Sunspot Number from 1700 to 1825. (Bottom panel) Sunspot group counts from 1 December 1769 to 28 February 1770. Blue, red, and green dots depict the sunspot observations made by Lalande, Horrebow, and Staudach, respectively. The vertical red and black lines represent the date of occurrence (1770 January 18) of the aurora event studied in the present work. Note that dates are in decimal numbers, not in "year-month" mode.

**Acknowledgments**

This research was supported by the Economy and Infrastructure Board of the Junta de Extremadura, and partially funded by FEDER - Junta de Extremadura (research group grant GR15137 and project IB16127) and by Spain's Ministerio de Economía y Competitividad (AYA2014-57556-P and CGL2017-87917-P). The authors are grateful for having benefited from participation in ISSI workshops. The authors acknowledge the helpful discussion with H. Hayakawa about calculations of the geomagnetic latitudes. We also appreciate comments made by D. Willis and the other anonymous referee in order to improve this manuscript.

**Disclosure of Potential Conflicts of Interest**

The authors declare that they have no conflicts of interest.